\documentclass[aps,
longbibliography,notitlepage,reprint,superscriptaddress]{revtex4-2}
\usepackage[colorlinks,linkcolor=blue,citecolor=blue,urlcolor=red]{hyperref}
\usepackage{soul}
\usepackage{changes}
\usepackage{float}
\usepackage{CJK}
\usepackage{graphicx}
\usepackage{amsmath,amssymb,amsfonts}
\usepackage{siunitx}
\usepackage{soul}
\usepackage{chngcntr}
\usepackage{chemformula}
\usepackage{MnSymbol}
\usepackage{dsfont}	
\usepackage{appendix}
\usepackage{lineno}

\newcommand{\HU}{John A. Paulson School of Engineering and Applied Sciences, Harvard University, Cambridge, MA 02138, USA}
\newcommand{\QSE}{Quantum Science and Engineering, Harvard University, Cambridge, MA 02138, USA}
\newcommand{\UOA}{Department of Physics, University of Auckland, Auckland, New Zealand}
\newcommand{\DODD}{The Dodd-Walls Centre for Photonic and Quantum Technologies, Dunedin, New Zealand}

\begin{document}

\author{Yunxiang Song}
\thanks{These authors contributed equally}
\affiliation{\HU}
\affiliation{\QSE}
\author{Jinsheng Lu}
\thanks{These authors contributed equally}
\affiliation{\HU}
\author{Xinrui Zhu}
\affiliation{\HU}
\author{Danxian Liu}
\affiliation{\HU}
\author{Zongda Li}
\affiliation{\UOA}
\affiliation{\DODD}
\author{Pawan Ratra}
\affiliation{\HU}
\author{Norman Lippok}
\affiliation{\HU}
\author{Miro Erkintalo}
\affiliation{\UOA}
\affiliation{\DODD}
\author{Federico Capasso}
\email{capasso@seas.harvard.edu}
\affiliation{\HU}
\author{Marko Lončar}
\email{loncar@g.harvard.edu}
\affiliation{\HU}

\title{Raman suppression in nanophotonics enabled by multimode spectral filtering}

\begin{abstract}
Miniaturized photonic cavities generating nonlinear optical states of light are central to telecommunications and metrology applications. The emergence of such states is primarily underpinned by the ubiquitous Kerr nonlinearity that is present in all media. However, stimulated Raman scattering (SRS), an additional process inherent to many materials, has been shown to critically hinder the states’ formation, imposing fundamental constraints on the choice of photonic platforms. Here, we introduce a novel strategy for the suppression of SRS in nanophotonic devices, adaptable to diverse Raman spectral responses. This is achieved by controlling the coupling and loss among multiple transverse spatial modes of the system, tailored across ultrabroad spectral bandwidths. Specifically, we combine nanometrically-corrugated Bragg gratings and tapered waveguides that, together enable co-directional multimode coupling and mode-selective filtering. We use lithium niobate as an exemplary Raman-active material to realize the concept, and we demonstrate the robust generation of two distinct Kerr nonlinear states (corresponding to coherent optical frequency combs) using the fabricated devices. The simplicity and generality of the concept suggest wide applicability to classical and quantum light generation on many technologically-relevant platforms nominally plagued by SRS (e.g., silicon and diamond photonics). More broadly, our multimode spectral shaping and filtering concept opens a path forward for highly-structured, wavelength-specific losses in nanophotonic waveguides and cavities, with potential applications in ultrafast and nonlinear integrated photonics.

\end{abstract}

\maketitle

Stimulated Raman scattering (SRS) in low-loss optical fiber has long enabled broadband lasing, amplification, and frequency conversion schemes \cite{stolen1972raman, stolen1973raman, namiki2001ultrabroad}, and remains central to long-haul communications as well as ultrafast and nonlinear optics \cite{islam2002raman,rong2005continuous, dudley2006supercontinuum, li2024ultrashort}. At the same time, SRS can be detrimental in frontier applications, where it causes nonlinear crosstalk between wavelength-division multiplexing channels or limits power scaling in fiber lasers and amplifiers, necessitating the development of techniques for its suppression \cite{vengsarkar2002long, nodop2010suppression, essiambre2010capacity}. As photonic platforms move beyond fiber-based systems, similar considerations arise in integrated architectures, where multiple nonlinearities, including SRS, can coexist and compete.

Integrated nanophotonics is one such platform, poised to transform modern communications \cite{marin2017microresonator, jorgensen2022petabit, yang2022multi, rizzo2023massively, corcoran2025optical}, metrology \cite{newman2019architecture, obrzud2019microphotonic, suh2019searching, kudelin2024photonic, sun2024integrated, zhao2024all, moille2025versatile,jin2025microresonator}, and sensing \cite{suh2016microresonator, riemensberger2020massively, chen2023breaking, chen2025single, zhang2022large, tang2023single} technologies. Owing to strong field confinement in wavelength-scale structures, these systems exhibit strong optical nonlinearities, primarily through the Kerr effect, at low input powers. This has enabled coherent light generation spanning octave spectral bandwidths \cite{del2007optical, sayson2019octave, lu2020chip,black2022optical}, squeezed light sources compatible with large-scale fabrication \cite{yang2021squeezed, shen2025strong, ulanov2025quadrature}, and optical microcombs for petabit-per-second data transmission \cite{marin2017microresonator, jorgensen2022petabit, yang2022multi, rizzo2023massively, corcoran2025optical} and precision microwave synthesis \cite{kudelin2024photonic, sun2024integrated, zhao2024all, moille2025versatile,jin2025microresonator}.

A wide range of materials have been explored for realizing Kerr-nonlinear nanophotonic devices \cite{hausmann2014diamond, griffith2015silicon, pu2016efficient, xuan2016high, li2017stably, wilson2020integrated, chang2020ultra, lukin20204h, liu2020photonic, jung2021tantala, liu2021aluminum, weng2021directly, wang2024octave, song2024octave}. In several key platforms, however, including lithium niobate and silicon, the desired pure Kerr dynamics is disrupted by SRS, much as in optical fiber. In nanophotonic cavities, SRS frequency downshifts pump photons via coupling to vibrational modes (a.k.a. Stokes shift), thereby clamping the intracavity pump power and significantly altering the intracavity field \cite{boyd2008nonlinear}. Crystalline materials such as lithium niobate and silicon feature characteristically sharp and strong Raman gain spectra, which can lead to very low-threshold SRS, particularly in high-Q cavities that endow efficient resonant enhancement for the pump and Stokes fields. Without intentional design, the low-threshold SRS can deteriorate or entirely inhibit Kerr nonlinear processes in such cavities, limiting access to a broad class of nonlinear light sources. Realizing such sources would, however, substantially extend the functionality of ultrafast electro-optic and solid-state quantum photonic platforms based on thin-film integration of crystalline materials \cite{rong2005continuous, hausmann2014diamond, lukin20204h, wang2018integrated, wang2024lithium}.

Recognized as a central challenge, SRS suppression on-chip has been pursued through a variety of approaches, including engineering of cavity sizes \cite{okawachi2017competition}, control of cavity-waveguide coupling rates \cite{song2024octave, he2023high, gong2020photonic, lv2025broadband}, and exploiting anisotropic Raman responses \cite{wang2024lithium, song2025stable, nie2025soliton, song2026high}. While powerful, these approaches face several limitations. Reducing the cavity size, i.e., increasing the cavity free spectral range (FSR), is utilized in cases of narrowband Raman gain, but still suffers from low yield and is unsuitable for applications that require small FSRs. Specialized cavity-waveguide couplers significantly distort the extracted nonlinear state, reducing the fidelity of the output relative to the intracavity field. Finally, methods relying on anisotropic Raman responses are highly material-specific and therefore lack applicability across material platforms. Given these limitations, a universal and platform-agnostic approach to SRS suppression is required to broaden the material scope of classical and quantum Kerr nonlinear nanophotonics. 

In this work, we introduce a novel design strategy that uses wavelength-dependent multimode control and mode-selective filtering (collectively termed multimode spectral filtering) to suppress SRS. The design incorporates long-period and nanometric Bragg gratings that couple co-propagating transverse spatial modes of the waveguide, followed by linear tapers that scatter higher-order modes. The core principle is to selectively convert intracavity signals, within targeted spectral ranges, from the low-loss fundamental spatial mode to a lossy higher-order spatial mode, by using only basic geometries (e.g., gratings and tapers) that exist on any integrated photonic platform. By structuring the total mode conversion profile (e.g., strength, wavelength, and bandwidth) based on grating design, SRS can be efficiently suppressed once this profile overlaps with the Raman gain spectrum. For proof-of-concept demonstration, we fabricate devices on thin-film lithium niobate (TFLN), a paradigmatic Raman-active material \cite{yu2020raman}. We show that a judiciously designed multimode spectral filter introduces negligible perturbations to the spectral range of interest, while successfully suppressing SRS in cavities of all dispersion conditions. Using these cavities, we demonstrate the generation of coherent Kerr frequency combs in anomalous and normal dispersion regimes, corresponding to two distinct nonlinear optical states also known as dissipative Kerr solitons \cite{leo2010temporal, chembo2013spatiotemporal, herr2014temporal, kippenberg2018dissipative} and normal dispersion localized structures \cite{huang2015mode, xue2015mode, lobanov2015frequency, helgason2021dissipative, liu2022stimulated}.

\begin{figure*}[t]
  	\includegraphics[width = 0.95\textwidth, page = 1]{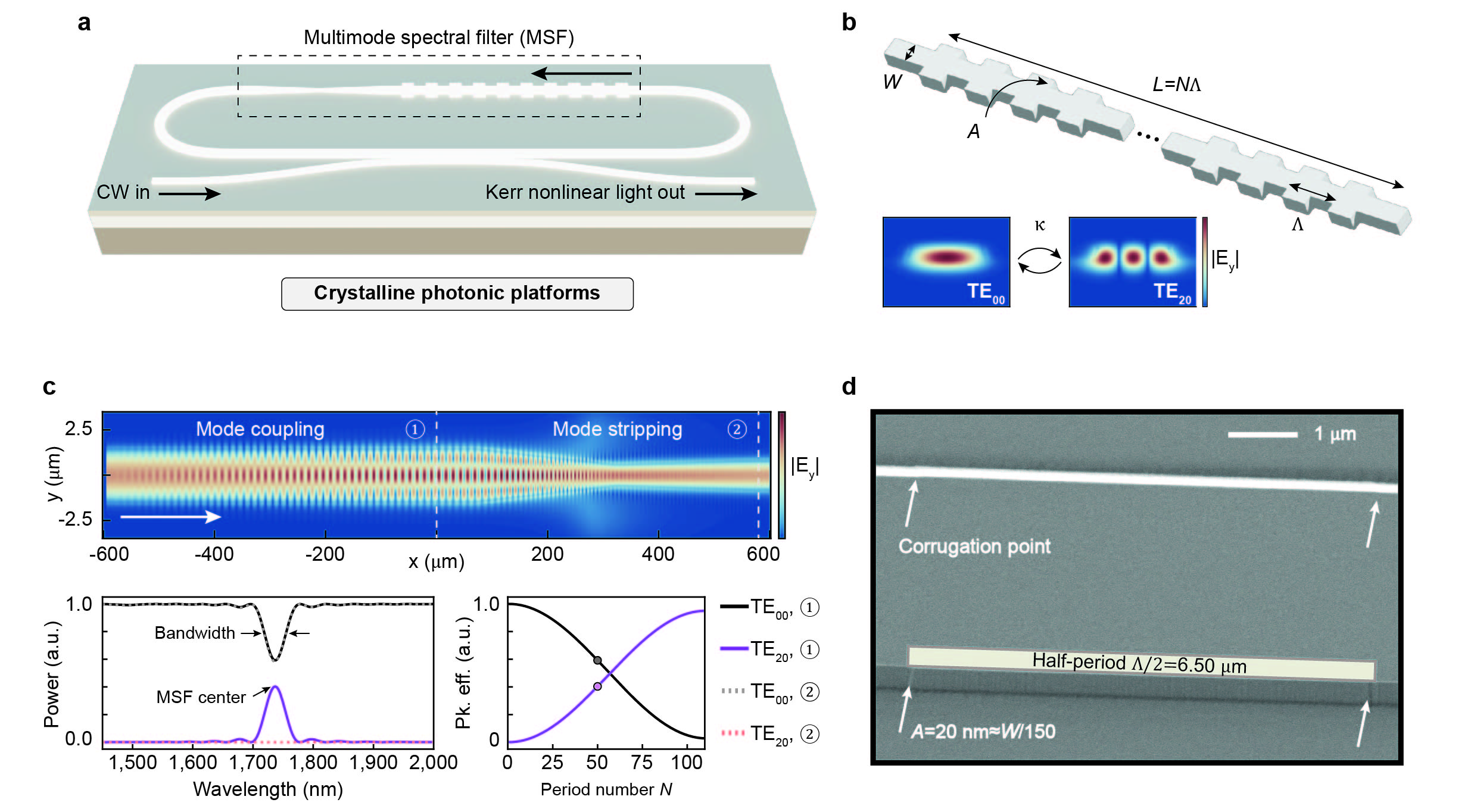}	\caption{\textbf{Concept.} \textbf{a}, Schematic illustration of the cavity design. A spectrally tailored multimode spectral filter (MSF, black dashed box) is embedded within a racetrack cavity. Under continuous-wave (CW) driving of the cavity, SRS is suppressed and Kerr nonlinear optical states may be generated. \textbf{b}, Schematic illustration of the MSF grating component. Design variables describing our single-period implementation are listed as follows: waveguide width ($W$), grating corrugation depth ($A$), grating period ($\Lambda$), grating number ($N$), and grating length ($L$). In our implementation, we selected a symmetric-rectangular grating corrugation pattern, with suitable values of $W$ and $\Lambda$, such that the co-propagating fundamental quasi-transverse-electric (TE$_{00}$) and second-order TE$_{20}$ transverse spatial modes are phase-matched at the Bragg wavelength (MSF center wavelength, $\lambda_0$). The intermodal coupling strength is denoted by $\kappa$. In \textbf{a} and \textbf{b}, the schematized sidewall corrugation amplitude is much exaggerated for visualization purposes. \textbf{c}, Electric field distribution across the MSF center-plane, extracted from finite-difference time-domain simulations considering an injected TE$_{00}$ mode at $\lambda_0$ (top). White arrow indicates the direction of light propagation and corresponds to the black arrow directions in \textbf{a}. End of the mode coupling region (constituting the grating component) is labeled by “1”, and end of the mode stripping region (constituting the taper) is labeled by “2”. Spectral response of the MSF (bottom left) represented by total power in the TE$_{00}$ and TE$_{20}$ modes at points 1 (solid lines) and 2 (dashed lines). A sinc-like response with roughly 40\% peak mode conversion efficiency is illustrated, given grating parameters of $W=2.93$ $\mu$m, $A=20$ nm, $\Lambda=13.0$ $\mu$m, and $N=50$. The tapered waveguide completes the mode conversion to loss transduction by linearly reducing in top width from $W$ to 1.13 $\mu$m, followed by an increase to 1.60 $\mu$m, over 300 $\mu$m propagation lengths, respectively. By varying $N$, the MSF grating component may operate at variable mode conversion peak efficiency from 0 to 1 (bottom right), and thus the MSF may operate at variable TE$_{00}$ attenuation factor $\alpha(\lambda_0)$ from 0 to arbitrarily large. The $N=50$ case corresponding to other subpanels within \textbf{c} is marked by the circles. \textbf{d}, Scanning electron microscope image of the MSF grating component. A half-period of $\Lambda/2=6.50$ $\mu$m is marked by the rectangle for clarity, and the nanometric corrugations on the waveguide sidewalls ($\approx1/150^{\text{th}}$ of the waveguide width) are pointed out by white arrows.}
	\label{fig:fig1}
\end{figure*}

\begin{figure*}[t]
    \includegraphics[width = 0.95\textwidth, page = 1]{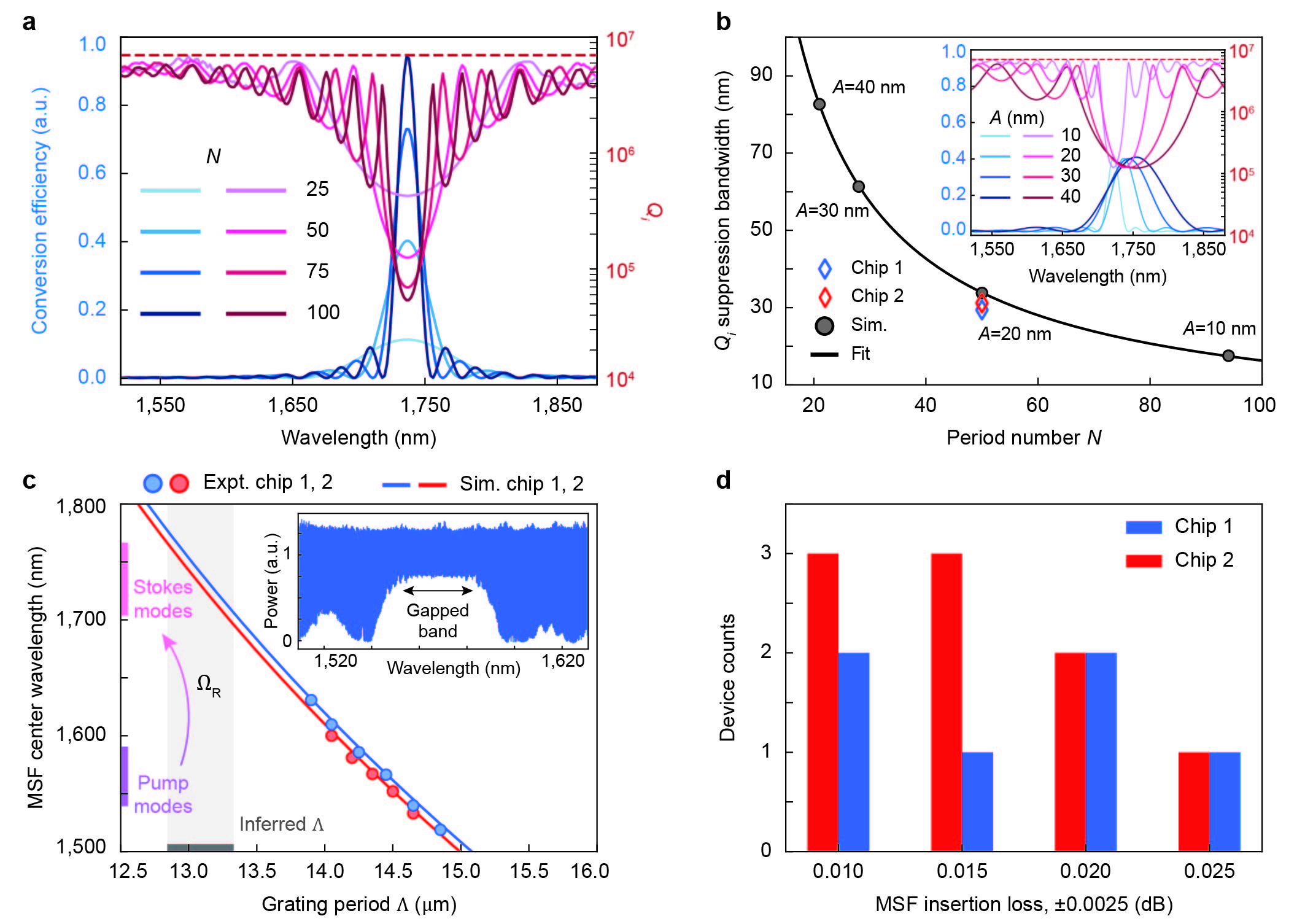}  
    \caption{\textbf{Intracavity multimode spectral filtering.} \textbf{a}, Simulated spectral response of the MSF grating component (blue curves) and the resulting frequency dependence of the cavity’s intrinsic quality factor $Q_i$ (red curves), as the grating number $N$ is varied. The MSF grating period is $\Lambda=13.0$ $\mu$m and the center wavelength is $\lambda_0=1,736$ nm, anticipating an $\approx18.8$ THz Stokes shift from C-band pump wavelengths. The red dashed line marks a frequency-independent reference intrinsic quality factor, $Q_{i,\text{ref}}$, of 7 million, achievable in a realistic anomalous-dispersion reference cavity without the MSF grating component, which is used as the baseline for computing the red curves. \textbf{b}, Measured (diamonds) and simulated (circles) $Q_i$ bandwidths as the grating corrugation depth $A$ is varied. For each $A$, the MSF is designed for roughly 40\% multimode conversion efficiency and $Q_i(\lambda_0)\approx0.1$ million (inset). The black line fits an inverse relationship between bandwidth and $N$ for constant conversion efficiency, according to MSF theory (see Supplementary information). The measured bandwidths are extracted from cavity transmission measurements. The blue (red) markers indicate bandwidths averaged over six devices on chip 1 (2). These averages are 31.16 and 29.40 nm, with standard deviations (not shown) of 1.39 and 0.68 nm, respectively. We note that the blue and red coloring scheme is used in subsequent panels as well, where the two chips are distinguished by anomalous dispersion and normal dispersion cavity designs. \textbf{c}, Measured (circles) and simulated (curves) MSF grating period $\Lambda$ vs. $\lambda_0$. The measured $\lambda_0$ are extracted from cavity transmission measurements within the C+L band (example shown in inset), and the simulated $\lambda_0$ are extracted from finite-difference eigenmode simulations based on as fabricated waveguide parameters. Candidate pump modes (within erbium-doped fiber amplifier bandwidth) are demarcated by the purple line (1,540 to 1,590 nm), corresponding Stokes-shifted Raman modes by the pink line (1,704 to 1,766 nm), and the inferred $\Lambda$ for compatible SRS suppression by the gray line (13.3 $\mu$m to 12.8 $\mu$m). The gray region serves as a guide to the eye. \textbf{d}, Measured MSF insertion losses for ten (six) devices on chip 1 (2), extracted from average $Q_i$ comparisons of cavity modes in the C+L band, between each device and its structurally similar reference cavity that lacks the MSF grating component. The histogram bins have centers as labeled and spans of $\pm$0.0025 dB.}
  \label{fig:fig2}
\end{figure*}

\section{Transmissive multimode spectral filter concept}

Figure 1a shows a high-level illustration of our concept, which comprises a multimode spectral filter (MSF) integrated within a racetrack cavity. The MSF realizes spectrally selective loss through partial coupling of the cavity’s fundamental quasi-transverse-electric (TE$_{00}$) mode to a higher-order transverse spatial mode using a grating, followed by scattering into a continuum of slab modes via a tapered waveguide. This scattering prevents recirculation or back-conversion of the higher-order mode; thus, from the perspective of the TE$_{00}$ mode, the combined process can be treated as wavelength-dependent loss.

In its simplest form [Fig. 1b], the MSF grating component consists of a single spatial period $\Lambda=\lambda_0/\Delta n_{\text{eff}}(\lambda_0)$, where $\lambda_0$ denotes the target center wavelength for mode conversion (Raman suppression) and $\Delta n_{\text{eff}}(\lambda_0)$ is the effective index difference between the coupled transverse spatial modes (e.g., TE$_{00}$ and TE$_{20}$), evaluated at $\lambda_0$. In addition to the center wavelength, the overall spectral response (e.g., peak conversion efficiency and bandwidth) is further controlled by waveguide width $W$, grating corrugation depth $A$, and grating period number $N$. Broadly, these parameters determine the effective index, $n_{\text{eff}}$, of the confined mode families, the intermodal coupling per unit length, and the total coupling length, respectively, each to be considered with greater detail in the following section. To translate wavelength-specific mode conversion into loss, it is sufficient to use a taper that does not confine the higher-order mode but still maintains low loss for the fundamental mode.

Figure 1c visualizes the operation of the MSF, showing the evolution of the electric field amplitude when an initially pure TE$_{00}$ mode is launched through the system. As the TE$_{00}$ mode propagates, it is gradually converted into the TE$_{20}$ mode, as evidenced by the emergence of a characteristic three-lobed field profile that interferes with the fundamental mode. The interference vanishes after the taper, indicating complete scattering of the TE$_{20}$ mode in a single pass. The remaining field exclusively occupies the TE$_{00}$ mode but with a reduced power relative to the input, corresponding to a filter extinction ratio of roughly 40\% (2.2 dB). The MSF exhibits a sinc-shaped spectral response with minimal parasitic filtering outside the intended stop band. While the MSF design is general, here we selected $\Lambda=$13.0 $\mu$m which corresponds to a center wavelength of $\lambda_0=1,736$ nm. This is a realistic grating period for suppressing Stokes modes in a telecom C-band-pumped lithium niobate cavity, considered in our experiments to be described in the subsequent sections.

Conceptually, near-ideal Raman suppression requires that the suppression mechanism introduces minimal additional loss channels within the spectral range of interest (e.g., bandwidth of the Kerr nonlinear optical state). This consideration motivates the use of co-propagating mode coupling (long-period $\Lambda$), rather than counter-propagating mode coupling (sub-wavelength $\Lambda$) as in many photonic-crystal-type cavities \cite{arbabi2011realization, lu2022high, moille2023fourier, lucas2023tailoring, ulanov2024synthetic, chen2024integrated, liu2024integrated}. Owing to the large phase mismatch between counter-propagating modes, the resulting conversion bandwidth (stop band) is typically very narrow unless high-index-contrast designs are employed, in turn introducing substantial excess loss and fabrication complexity. In addition, reflective coupling processes to undesired mode families are poorly separated in wavelength, which will result in spuriously suppressed spectral regions. By contrast, co-propagating mode conversion involves only a small phase mismatch and yields well-separated conversion bands, enabling shallow gratings to produce nearly lossless, broadband spectral filtering. As shown in Fig. 1d, a fully functional MSF based on the design in Fig. 1c displays only nanometric corrugations; since the grating component is nearly indistinguishable from a uniform waveguide, TE$_{00}$ scattering losses are negligible.

\section{Implementation on lithium niobate for Raman suppression}

To confirm the viability and further elaborate upon the concept described above, we conducted a detailed parametric study, combining numerical simulations with experiments. As a test case, we implement the cavity design on z-cut TFLN, whose TE$_{00}$ mode nominally exhibits ultralow-threshold Raman lasing. In this platform, SRS is characterized by a Stokes shift of $\Omega_R\approx18.8$ THz and a Stokes bandwidth of $\Gamma_R\approx560$ GHz \cite{kaminow1967quantitative}. Following illustrations in Fig. 1, we realize Raman-suppressed cavities using intracavity MSFs that couple the TE$_{00}$ and TE$_{20}$ mode pair.

A key metric of the MSFs is their Raman suppression efficiency (i.e., filter extinction ratio), in our scheme directly related to the wavelength-dependent multimode conversion efficiency $\eta(\lambda)$. Expanding on the simulations in Fig. 1c, we consider a CW drive with pump wavelength $\lambda_p\approx1,565$ nm (C-band); the CW Raman response thus peaks at $\lambda_S\approx1,736$ nm, spanning a 3-dB bandwidth of $\Delta\lambda\approx5.6$ nm. Taking the MSF grating component to be single-period with $\Lambda=13.0$ $\mu$m, $W=2.93$ $\mu$m, and $A=20$ nm, full-wave simulations show that the peak conversion efficiency $\eta(\lambda_0\approx\lambda_S)$ varies from 11\% to 94\%, as the number of grooves, $N$, is swept from 25 to 100 [Fig. 2a]. When the MSF is embedded within a photonic cavity, the intrinsic quality factors, $Q_i(\lambda)$, of the cavity modes are related to $\eta(\lambda)$ by

\begin{equation}
Q_i(\lambda)=\frac{c}{\lambda\cdot f_{\text{FSR}}\cdot\big( 1-(1-\alpha_0)\cdot(1-\eta) \big)},
\end{equation}
where $c$ is the speed of light, $f_\text{FSR}$ is the FSR, $\alpha_0=c\cdot(\lambda\cdot f_{\text{FSR}}\cdot Q_{i,\text{ref}})^{-1}$ is the grating-free round-trip power loss, and $Q_{i,\text{ref}}$ is a typical grating-free (reference) intrinsic quality factor. Importantly, our simulations confirm that, in the presence of an MSF, the intrinsic quality factor at the pump wavelength, $Q_i(\lambda_p)$, can remain in the ultralow-loss regime, experiencing only less than a factor-of-two reduction relative to the unfiltered value $Q_{i,\text{ref}}$, while the quality factor at the target wavelength, $Q_i(\lambda_0)$, is reduced by more than a factor of fifty for $N>50$. This contrast between $Q_i(\lambda_p)$ and $Q_i(\lambda_0)$ is central to selectively suppressing SRS and, when maintained over a sufficiently broad bandwidth, renders the design tolerant to misalignment between the target ($\lambda_0$) and Stokes ($\lambda_S$) wavelengths as well (thus implying pump wavelength flexibility). We define the Raman suppression bandwidth as the wavelength range encompassed by the 3-dB suppression points around $Q_i(\lambda_0)$, and practical considerations for designing this bandwidth are further detailed below.

For a fixed $Q_i$ contrast, the bandwidth of co-propagating multimode conversion (equivalently, Raman suppression bandwidth) is primarily governed by the coupling strength per unit length, linked to the corrugation depth $A$. In Fig. 2b, we simulate MSF designs that provide identical Raman suppression efficiency as $A$ varies from 10 to 40 nm, and we plot their corresponding bandwidths. Not surprisingly, the period number $N$ (related to coupling length) must increase for smaller $A$ to achieve the same suppressed $Q_i(\lambda_0)$, and the associated bandwidths are inversely proportional to $N$ (see Supplementary information). We observe clear design trade-offs: a broader suppression bandwidth offers greater pump wavelength flexibility, at the expense of greater perturbations beyond the region of interest introduced by far-reaching sidelobes of the filter function. Conversely, an overly narrow suppression bandwidth (relative to the Stokes bandwidth $\Gamma_R$) may fail to suppress all cavity modes capable of Raman lasing for a given pump wavelength, $\lambda_p$, though it minimally perturbs $Q_i$ elsewhere. Further, we note that bandwidth engineering through $A$ must be accompanied by adjustments to $W$, since changing $A$ alone changes the average effective index of the grating and hence $\lambda_0$ shifts slightly [Fig. 2b, inset].

\begin{figure*}[t]
    \includegraphics[width = 0.95\textwidth, page = 1]{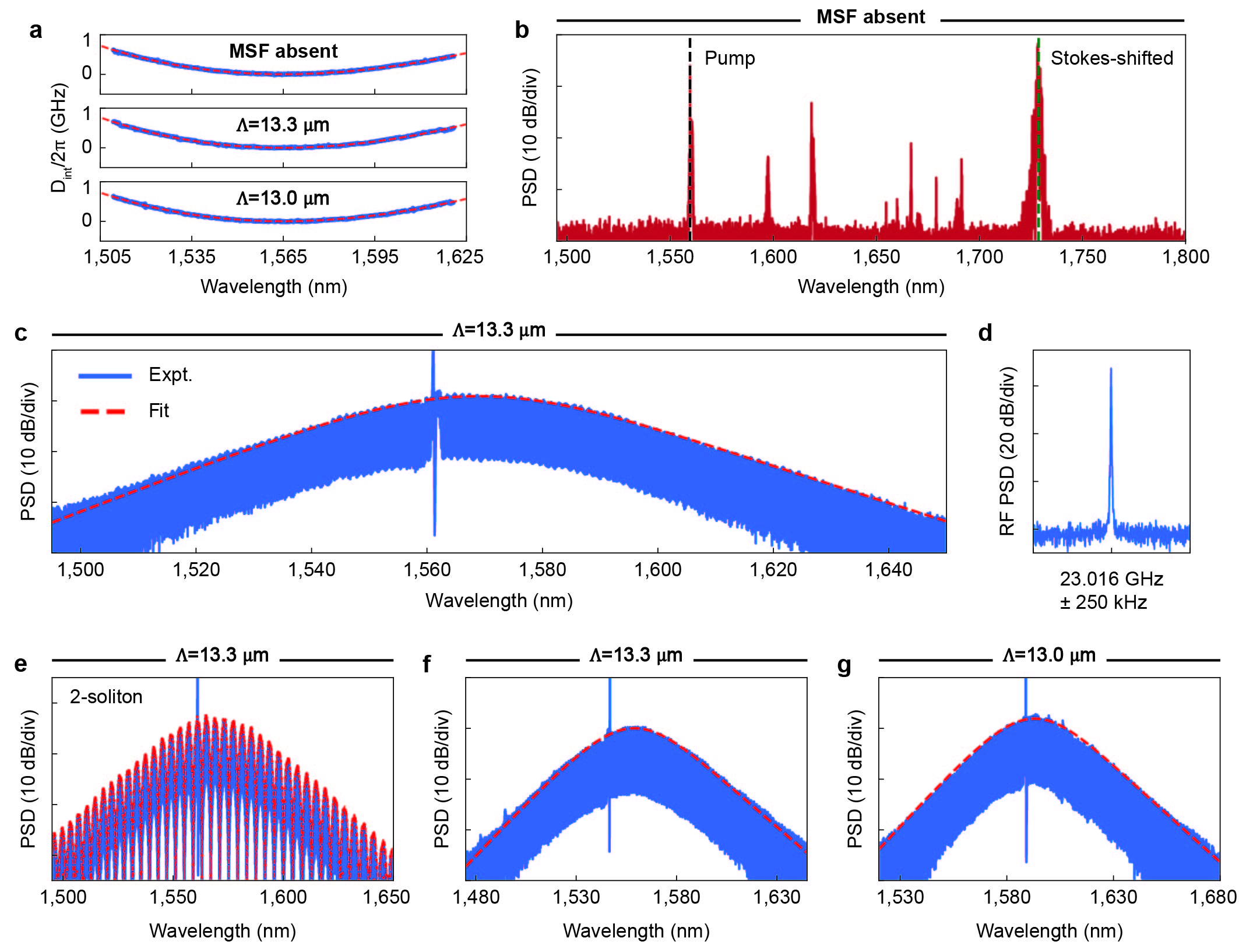}  
    \caption{\textbf{Dissipative Kerr solitons in anomalous dispersion cavities.} \textbf{a}, Integrated dispersions ($D_\text{int}$) of three devices: reference cavity without the MSF grating component (top), cavity with $\Lambda=13.3$ $\mu$m MSF (middle), and cavity with $\Lambda=13.0$ $\mu$m MSF (bottom). Anomalous group-velocity dispersions are achieved. Blue dots are extracted from cavity transmission measurements, and red dashed curves are derived from quartic polynomial fits in the mode number space. \textbf{b}, Raman lasing and four-wave-mixing spectrum from the reference cavity, without an intracavity MSF, under an on-chip CW drive with roughly 350 mW and $\lambda_p=1,559.66$ nm. Black (green) dashed line marks the pump (Stokes-shifted) light. The many lines in the Stokes-shifted band are due to multimode Raman lasing, where many cavity modes satisfy the lasing condition. Additional spectral lines between the prominent pump and Stokes lines are due to four-wave mixing. \textbf{c}, Single soliton spectrum from the $\Lambda=13.3$ $\mu$m-MSF cavity, with pump wavelength in the C-band ($\lambda_p=1,561.05$ nm) and identical pump power as in \textbf{b}. Red dashed curves here, and in panels \textbf{f} and \textbf{g}, indicate fits to sech$^2$ profiles. \textbf{d}, Repetition-rate beatnote of the soliton in \textbf{c}. The measurement span is 500 kHz and resolution bandwidth is 250 Hz. \textbf{e}, Multi-soliton spectrum obtained by driving the same cavity and resonance as in \textbf{c}. Red dashed curve indicates a fit to a two-soliton state spectrum. \textbf{f}, Single soliton spectrum obtained by driving the same cavity as in \textbf{c}, but at a different resonance ($\lambda_p=1,546.83$ nm) that still satisfies the design condition. \textbf{g}, Single soliton spectrum from the $\Lambda=13.0$ $\mu$m-MSF cavity, with pump wavelength ($\lambda_p=1,588.98$ nm) in the L-band.}
  \label{fig:fig3}
\end{figure*}

\begin{figure}[!htbp]
    \includegraphics[width = 0.475\textwidth, page = 1]{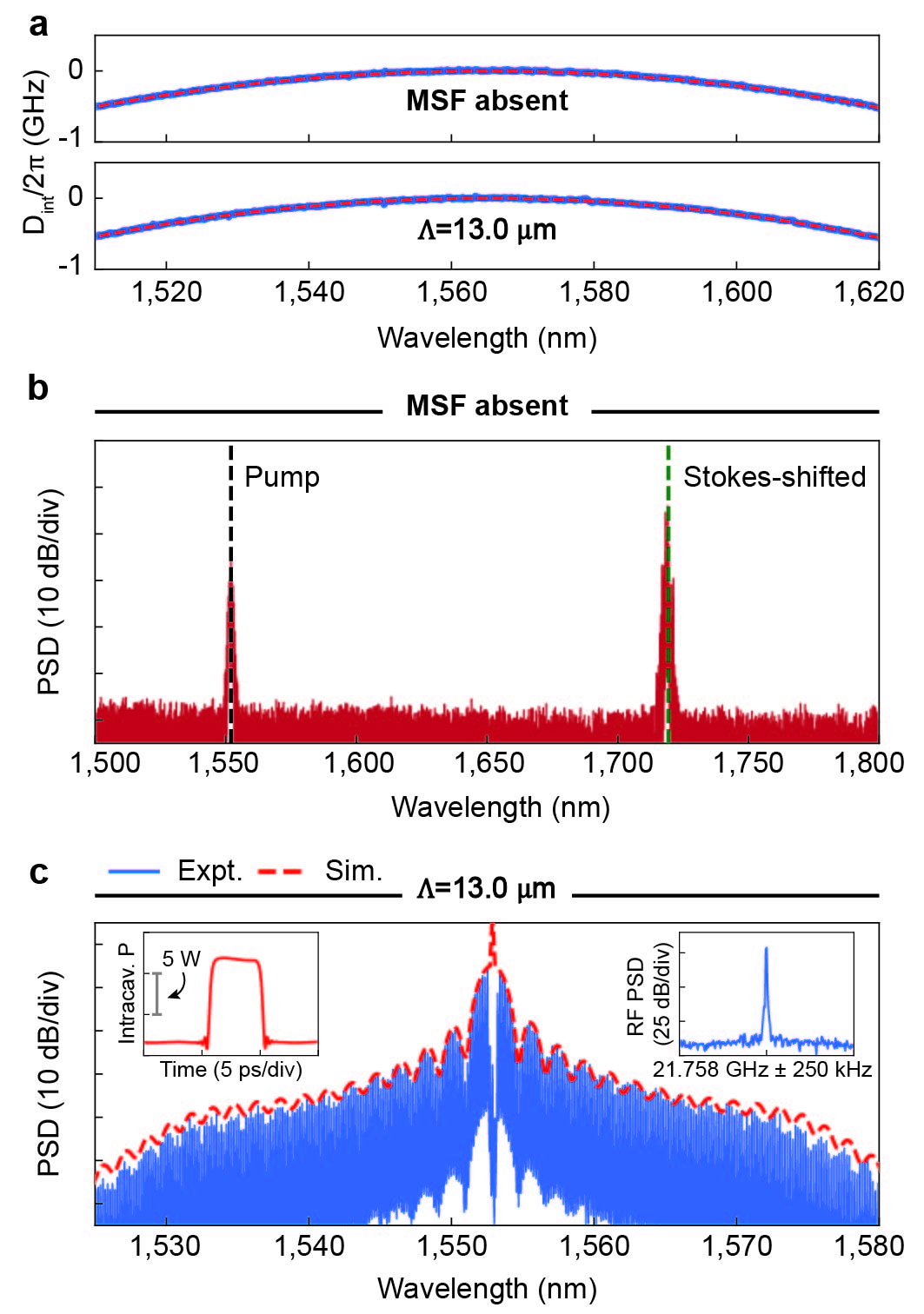}  
    \caption{\textbf{Coherent normal dispersion microcombs.} \textbf{a}, Integrated dispersions ($D_\text{int}$) of two devices: reference cavity without the MSF grating component (top) and cavity with $\Lambda=13.0$ $\mu$m MSF (bottom). Normal group-velocity dispersions are achieved. Blue dots are extracted from cavity transmission measurements, and red dashed curves are derived from quartic polynomial fits in the mode number space. \textbf{b}, Raman lasing and four-wave-mixing spectrum from the reference cavity, without an intracavity MSF, under an on-chip CW drive with roughly 150 mW and $\lambda_p=1,552.05$ nm. Black (green) dashed line marks the pump (Stokes-shifted) light. \textbf{c}, Normal dispersion microcomb spectrum from the $\Lambda=13.0$ $\mu$m-MSF cavity, with pump wavelength in the C-band ($\lambda_p=1,552.81$ nm) and identical pump power as in \textbf{b}. Red dashed curve overlays a simulated spectrum (Lugiato-Lefever equation), in good agreement with the experimental data (blue lines). Left inset shows the simulated temporal waveform, which is a characteristically rectangular pulse corresponding to the sinc$^2$-like oscillations in the spectral envelope. Right inset shows the repetition-rate beatnote, where the measurement span is 500 kHz and the resolution bandwidth is 250 Hz.}
  \label{fig:fig5}
\end{figure}

Considering the particular SRS response of z-cut TFLN, the above design rules allow us to identify an MSF grating design with a Raman suppression strength corresponding to roughly 40\% round-trip power loss and suppression bandwidth of $\approx30$ nm ($W=2.93$ $\mu$m, $A=20$ nm, $N=50$) as a robust operating point that favors Kerr dynamics over competing SRS. The Raman suppression strength and bandwidth are experimentally verified in our fabricated TFLN cavities (Supplementary information and Fig. 2b, respectively).

To determine $\Lambda$ such that $\lambda_0\approx\lambda_S$ is satisfied, we experimentally sweep $\Lambda$ and measure the corresponding $\lambda_0$ over the range 1,510-1,630 nm (set by the tuning range of our tabletop external-cavity diode laser). Figure 2c plots $\Lambda$ vs. $\lambda_0$ for two chips, where “chip 1”  and “chip 2” designate chips with exclusively anomalous and normal dispersion cavity designs, respectively. Notably, the MSFs in all cavities share identical parameters, and the data from both chips show excellent agreement despite the differing dispersion conditions. Residual offsets arise from fabrication variation, as the grating width $W$ is weakly coupled to the wet etching temperature and duration (see Supplementary information). Accounting for these variations, index simulations accurately reproduce experimental trends [curves in Fig. 2c, differentiated by $W=2.94$ $\mu$m (blue) and $W=2.93$ $\mu$m (red)]. From this, we infer that grating periods $\Lambda$ between 12.8 and 13.3 $\mu$m are well-suited for pump wavelengths in the C- and L-bands.

By comparing the average $Q_i$ of cavity modes in the C- and L-bands for devices with and without the MSF grating component, we experimentally find that all the gratings introduce (unwanted) excess losses below 0.03 dB per pass [Fig. 2d]. This result confirms the low parasitic filtering and negligible scattering loss of the grating outside the intended filter stop band. As will be shown, this enables high-fidelity generation and extraction of Kerr nonlinear states from the cavities, consistent with established wisdom.

\section{Raman-free coherent Kerr microcombs}
The modeling and experiments reported above clearly demonstrate that our concept allows us to introduce wavelength-specific losses in photonic cavities. We now show that this ability indeed enables efficient suppression of SRS in TFLN, which in turn permits the generation of coherent Kerr microcombs in devices with overall anomalous and normal dispersions that would not be feasible in the absence of an MSF.

We first consider the generation of dissipative Kerr soliton frequency combs in devices featuring anomalous dispersion. In Fig. 3a, we compare the integrated dispersion ($D_\text{int}$) of a reference cavity without the MSF grating component and two cavities incorporating MSFs with $\Lambda=13.3$ $\mu$m and $\Lambda=13.0$ $\mu$m. The upward-facing parabolic shapes indicate anomalous group-velocity dispersion – a dispersion condition suitable for soliton generation in the absence of SRS dynamics – is maintained despite the embedded MSFs.

As expected, CW driving of the reference cavity (which does not include an MSF) leads to Raman lasing through SRS. This is confirmed by injecting roughly 350 mW of CW power at $\lambda_p=1,559.66$ nm into the cavity, yielding an output spectrum dominated by two strong lines at the pump and Stokes-shifted wavelengths [Fig. 3b]. In contrast, injecting similar CW power at $\lambda_p=1,561.05$ nm into the cavity with a $\Lambda=13.3$ $\mu$m-MSF yields the optical spectrum shown by the blue curve in Fig. 3c. The spectral envelope is well fitted by the sech$^2$ function [red curve, Fig. 3c], which is consistent with a single soliton state. Since the cavity FSR lies in the microwave regime ($\approx23.35$ GHz), the coherence of this state is directly evidenced by its strong and narrow microwave beatnote [Fig. 3d]. By slightly adjusting the pump detuning through $\lambda_p$, the same cavity mode supports multi-soliton states, for example a two-soliton state [Fig. 3e]. We find that, thanks to the bandwidth of Raman suppression, cavity modes across the entire C-band exhibit Raman-free Kerr dynamics. To show this, we demonstrate single-soliton generation at a pump wavelength detuned from the previous one by almost 15 nm ($\lambda_p=1,546.83$ nm, see Fig. 3f). In another cavity with a $\Lambda=13.0$ $\mu$m-MSF [Fig. 3g], the filter stop band is blue-shifted, and accordingly the single soliton pump band is shifted to the L-band ($\lambda_p=1,588.98$ nm).

In addition to soliton states in the anomalous dispersion regime, our concept can also enable the generation of microcombs in the normal dispersion regime. In Fig. 4a, we show that both a reference cavity and a cavity incorporating an MSF with $\Lambda=13.0$ $\mu$m feature similar $D_{\text{int}}$ curves that are downward-facing parabolas, indicating overall normal group-velocity dispersion. Aligned with our previous CW-driving experiments, the reference cavity exhibits strong Raman lasing [Fig. 4b] and the MSF-embedded cavity hosts coherent normal dispersion microcombs [Fig. 4c]. These combs exhibits sinc$^2$-like spectral lobes around the pump ($\lambda_p=1,552.81$ nm) and broad wings at both spectral edges, in good agreement with simulations from the Lugiato-Lefever equation that uses experimental parameters – these simulations also suggest that the comb states correspond to flat-top rectangular pulses in the time domain [Fig. 4c, left inset]. The coherence of the state is further confirmed by its strong and narrow repetition-rate beatnote [Fig. 4c, right inset].

\section{Discussion}

In conclusion, we presented a transmissive grating-based design for nanophotonic cavities that efficiently suppresses SRS through wavelength-selective loss engineering across multiple spatial modes. Importantly, the design approaches an ideal implementation: it minimally perturbs cavity modes outside the Raman gain spectrum while preserving ultralow loss over a broad spectral range. We further demonstrated that our design enables the robust generation of Kerr frequency combs in lithium niobate cavities featuring anomalous and normal group-velocity dispersions, highlighting our design’s compatibility with all nonlinear regimes.

Our results provide a universal solution to a long-standing challenge in integrated photonics: many material platforms exhibit deleterious Raman effects but lack a noninvasive strategy for suppressing them. Our design relies on fundamental photonic elements – shallow Bragg gratings and tapered waveguides – and is therefore readily transferrable across platforms, opening new application spaces. For example, silicon cavities have recently shown ultralow loss in the telecom/near-infrared wavelength bands \cite{burla2015ultra,zhang2020ultrahigh}; our design may enable silicon-photonic Kerr microcombs and squeezed light manufactured at foundry scales. Diamond \cite{hausmann2014diamond} and silicon carbide \cite{lukin20204h} photonics host native quantum defects, and their interaction with monolithic nonlinear light sources may be studied in SRS-suppressed resonators. Lithium niobate \cite{hu2025integrated} and tantalate \cite{wang2024lithium} cavities that feature strong electro-optic or second-order nonlinear-optical properties require robust Raman suppression, and our work may finally unlock high-power resonant electro-optic comb and harmonic generation, as well as the study of intracavity hybrid nonlinear interactions.

Lastly, it is worth noting that our devices physically resemble photonic crystal-type cavities that have recently attracted attention for dispersion engineering \cite{arbabi2011realization, lu2022high, moille2023fourier, lucas2023tailoring, ulanov2024synthetic, pimbi2026full} and vortex beam generation \cite{chen2024integrated, liu2024integrated}. Functionally, however, our approach benefits from an expanded design space enabled by mode-selective processes. While the modal degree of freedom on-chip has been widely explored in the context of photonic quantum computing \cite{mohanty2017quantum}, transceivers \cite{luo2014wdm, stern2015chip}, signal processing \cite{khaled2025fully}, and spectral shaping \cite{lu2025cascaded} applications, explicit mode conversion has only rarely been used in resonant configurations \cite{ginis2023resonators, tao2024versatile, jha2025efficient}. Extending this capability to resonant cavities is particularly promising, as modal coupling engineering can enhance nonlinear light–matter interactions or imprint nearly arbitrary loss spectra into the cavity, for example. Accordingly, our transmissive multimode converters also offer a promising route towards spatially multiplexed and cavity-based on-chip nonlinear light sources \cite{lucas2018spatial}.

\section*{Acknowledgments}
We thank Yan Yu (Caltech) for discussions. The device fabrication was performed at the Harvard University Center for Nanoscale Systems (CNS); a member of the National Nanotechnology Coordinated Infrastructure Network (NNCI), which is supported by the National Science Foundation under NSF award no. ECCS-2025158.

\section*{Contributions}
Y.S. conceived the idea for the project. Y.S. and J.L. designed the devices and performed the simulations. Y.S. and X.Z. fabricated the chips. Y.S. and J.L. performed proof-of-concept experiments. D.L. performed nonlinear experiments with Y.S. assisting. Y.S., D.L., Z.L., and J.L. interpreted and analyzed the data. P.R. and N.L. helped with the project. Y.S. wrote the manuscript with significant contributions from M.E., F.C., and M.L.; all other authors commented on the manuscript. M.E., F.C., and M.L. supervised the project.

\section*{Funding}
Naval Air Warfare Center Aircraft Division (N6833522C0413), Department of Defense (FA9453-23-C-A039), National Science Foundation (ECCS-2407727), Air Force Lifecycle Management Center (FA8702-15-D-0001), Air Force Office of Scientific Research (FA955025CB011), Department of the Navy (N6833525C0336), Naval Sea Systems Command (N0002425CT146), National Research Foundation of Korea (NRF-2022M3K4A1094782), Amazon Web Services (A60290), Multidisciplinary University Research Initiative from the Air Force Office of Scientific Research (FA9550-22-1-0307), and Marsden Fund of the Royal Society Te Apārangi of New Zealand.

\section*{Competing interests}
M.L. is involved in developing lithium niobate technologies at HyperLight Corporation. The authors declare no other competing interests.

\section*{Data availability}
All data needed to evaluate the conclusions in the paper are present in the paper and/or the Supplementary Information.




\bibliographystyle{apsrev4-2}
\bibliography{refs}

\end{document}